\documentclass[12pt]{article}
\usepackage{graphicx}

\begin{document}

\baselineskip=18pt plus 1pt minus 1pt
\begin{center} 

{\large \bf X(3): an exactly separable $\gamma$-rigid version of the X(5) 
critical point symmetry}

\bigskip\bigskip

{Dennis Bonatsos$^{a}$\footnote{e-mail: bonat@inp.demokritos.gr},
D. Lenis$^{a}$\footnote{e-mail: lenis@inp.demokritos.gr}, 
D. Petrellis$^{a}$\footnote{e-mail: petrellis@inp.demokritos.gr}, 
P. A. Terziev$^{b}$\footnote{e-mail: terziev@inrne.bas.bg} },
I. Yigitoglu$^{a,c}$\footnote{e-mail: yigitoglu@istanbul.edu.tr}
\bigskip

{$^{a}$ Institute of Nuclear Physics, N.C.S.R.
``Demokritos''}

{GR-15310 Aghia Paraskevi, Attiki, Greece}

{$^{b}$ Institute for Nuclear Research and Nuclear Energy, Bulgarian
Academy of Sciences }

{72 Tzarigrad Road, BG-1784 Sofia, Bulgaria}

{$^{c}$ Hasan Ali Yucel Faculty of Education, Istanbul University}

{TR-34470 Beyazit, Istanbul, Turkey} 

\end{center}

\bigskip\bigskip
\centerline{\bf Abstract} \medskip 

A $\gamma$-rigid version (with $\gamma=0$) of the X(5) critical point symmetry
is constructed. The model, to be called X(3) since it is proved to contain
three degrees of freedom, utilizes an infinite well potential, 
is based on exact separation of variables,  
and leads to parameter free (up to overall scale factors) predictions 
for spectra and $B(E2)$ transition rates, which are in good agreement with 
existing experimental data for $^{172}$Os and $^{186}$Pt. An unexpected 
similarity of the $\beta_1$-bands of the X(5) nuclei $^{150}$Nd, $^{152}$Sm,
$^{154}$Gd, and $^{156}$Dy to the X(3) predictions is observed.

\newpage 

\section{Introduction}

Critical point symmetries \cite{IacE5,IacX5}, describing nuclei at 
points of shape phase transitions between different limiting symmetries,
have recently attracted considerable attention, since they lead to 
parameter independent (up to overall scale factors) predictions which 
are found to be in good agreement with experiment 
\cite{CZ1,Clark1,CZ2,Clark2}. The X(5) 
critical point symmetry \cite{IacX5}, in particular, is supposed 
to correspond to the transition from vibrational [U(5)] to prolate axially 
symmetric [SU(3)] nuclei, materialized in the $N=90$ isotones 
$^{150}$Nd \cite{Kruecken}, $^{152}$Sm \cite{CZ2}, $^{154}$Gd 
\cite{Tonev,Dewald}, and $^{156}$Dy \cite{Dewald,CaprioDy}. 

On the other hand, it is known that in the framework of the nuclear collective
model \cite{Bohr}, which involves the collective variables $\beta$ and 
$\gamma$, interesting special cases occur by ``freezing'' the $\gamma$ variable
\cite{DC} to a constant value.

In the present work we constuct a version of the X(5) model in which 
the $\gamma$ variable is ``frozen'' to $\gamma=0$, instead of varying 
around the $\gamma=0$ value within a harmonic oscillator potential, as in
the X(5) case. It turns out that only three variables are involved 
in the present model, which is therefore called X(3). Exact separation 
of the $\beta$ variable from the angles is possible. Experimental 
realizations of X(3) appear to occur in $^{172}$Os and $^{186}$Pt, while an 
unexpected agreement of the $\beta_1$-bands of the X(5) nuclei $^{150}$Nd, 
$^{152}$Sm, $^{154}$Gd, and $^{156}$Dy to the X(3) predictions is observed. 

In Section 2 the X(3) model is constructed, while numerical results 
and comparisons to experiment are given in Section 3, and a discussion 
of the present results and plans for further work in Section 4. 

\section{The X(3) model}

In the collective model of Bohr \cite{Bohr} the classical expression of the 
kinetic energy corresponding to $\beta$ and $\gamma$ vibrations of the nuclear
surface plus rotation of the nucleus has the form \cite{Bohr,ST}
\begin{equation}\label{eq:e1}
T = \frac{1}{2}\sum_{k=1}^{3} {\cal J}_k\, \omega^{\prime2}_k +
 \frac{B}{2}\,(\dot{\beta}^2+\beta^2 \dot{\gamma}^2), 
\end{equation}
where $\beta$ and $\gamma$ are the usual collective variables, $B$ is the 
mass parameter, 
\begin{equation}\label{eq:e2}
{\cal J}_k = 4B\beta^2 \sin^2\bigl(\gamma - {\textstyle\frac{2}{3}}\pi k\bigr)
\end{equation}
are the three principal irrotational moments of inertia, 
and $\omega^\prime_k$ ($k=1$, 2, 3) are the components of the angular velocity
on the body-fixed $k$-axes,
which can be expressed in terms of the time derivatives of the Euler angles 
$\dot{\phi}, \dot{\theta}, \dot{\psi}$ \cite{ST,Zare}  
\begin{eqnarray}
\omega^\prime_1 &=& -\sin\theta \cos\psi\,\dot{\phi} + \sin\psi\,\dot{\theta}, 
\nonumber\\
\omega^\prime_2 &=& \sin\theta \sin\psi\,\dot{\phi} + \cos\psi\,\dot{\theta},\\
\omega^\prime_3 &=& \cos\theta\,\dot{\phi} + \dot{\psi}. \nonumber
\end{eqnarray}
Assuming the nucleus to be $\gamma$-rigid (i.e. $\dot{\gamma}=0$), as in the 
Davydov and Chaban approach \cite{DC}, and considering in particular 
the axially symmetric prolate case of $\gamma=0$, we see that  
the third irrotational moment of inertia ${\cal J}_3$ vanishes, while 
the other two become equal ${\cal J}_1 = {\cal J}_2 = 3B\beta^2$, the kinetic 
energy of Eq. (\ref{eq:e1}) reaching the form \cite{ST,Davydov} 
\begin{equation}\label{eq:e4}
T = \frac{1}{2} 3B\beta^2 (\omega^{\prime2}_1 + \omega^{\prime2}_2) 
+ \frac{B}{2}\,\dot{\beta}^2
= \frac{B}{2}\Bigl[3\beta^2(\sin^2\theta\,\dot{\phi}^2 + \dot{\theta}^2) 
+ \dot{\beta}^2 \Bigr].
\end{equation}
It is clear that in this case the motion is characterized by three degrees 
of freedom. Introducing the generalized coordinates $q_1=\phi$, 
$q_2=\theta$, and $q_3=\beta$, the kinetic energy becomes 
a quadratic form of the time derivatives of the generalized coordinates
\cite{ST,EG}
\begin{equation}\label{eq:e5}
T=\frac{B}{2}\sum_{i,j=1}^3 g_{ij}\; \dot{q}_i\dot{q}_j, 
\end{equation}
with the matrix $g_{ij}$ having a diagonal form
\begin{equation}\label{eq:e6} 
g_{ij} = \left(\begin{array}{ccc} 3\beta^2\sin^2\theta & 0 & 0 \\ 0 & 
3\beta^2 & 0 \\0 & 0 & 1 \end{array}\right). 
\end{equation}
(In the case of the full Bohr Hamiltonian \cite{Bohr} the square matrix 
$g_{ij}$ is 5-dimensional and non-diagonal \cite{ST,EG}.)
Following the general procedure of quantization in curvilinear coordinates
one obtains the Hamiltonian operator \cite{ST,EG}
\begin{equation}
H = -\frac{\hbar^2}{2B}\,\Delta + U(\beta)
= -\frac{\hbar^2}{2B}\Biggl[\frac{1}{\beta^2}
\frac{\partial}{\partial\beta}\beta^2\frac{\partial}{\partial\beta} +
\frac{1}{3\beta^2} \Delta _\Omega\Biggr] + U(\beta), 
\end{equation}
where $\Delta_\Omega$ is the angular part of the Laplace operator
\begin{equation}
\Delta_\Omega = \frac{1}{\sin\theta}\frac{\partial}{\partial\theta}\sin\theta
\frac{\partial}{\partial\theta} + \frac{1}{\sin^2\theta}
\frac{\partial^2}{\partial\phi^2}. 
\end{equation}
The Schr\"odinger equation can be solved by the factorization
\begin{equation} \label{eq:e8}
\Psi(\beta,\theta,\phi) = F(\beta)\, Y_{LM}(\theta,\phi), 
\end{equation}
where $Y_{LM}(\theta,\phi)$ are the spherical harmonics. Then 
the angular part leads to the equation 
\begin{equation}\label{eq:9}
-\Delta_\Omega Y_{LM}(\theta,\phi) = L(L+1)Y_{LM}(\theta,\phi), 
\end{equation}
where $L$ is the angular momentum quantum number, 
while for the radial part $F(\beta)$ one obtains
\begin{equation}\label{eq:e9}
\Biggl[\frac{1}{\beta^2}
\frac{d}{d\beta}\beta^2\frac{d}{d\beta}
-\frac{L(L+1)}{3\beta^2} + \frac{2B}{\hbar^2}\Bigl(E-U(\beta)\Bigr)\Biggr] 
F(\beta) = 0. 
\end{equation}
As in the case of X(5) \cite{IacX5}, the potential in $\beta$ is taken to be 
an infinite square well
\begin{equation}\label{eq:e10}
U(\beta) = \left\{ \begin{array}{lcl} 0, && 0\leq\beta\leq\beta_W \\
\infty, && \beta > \beta_W \end{array} \right. ,
\end{equation}
where $\beta_W$ is the width of the well. 
In this case $F(\beta)$ is a solution of the equation
\begin{equation}\label{eq:e11}
\Biggl[\frac{d^2}{d\beta^2} + \frac{2}{\beta}\frac{d}{d\beta}
 + \Biggl(k^2-\frac{L(L+1)}{3\beta^2}\Biggr)\Biggr] F(\beta) = 0
\end{equation}
in the interval $0\leq\beta\leq\beta_W$, 
where reduced energies $\varepsilon=k^2=2B E/\hbar^2$ \cite{IacX5}
have been introduced, while it vanishes outside.
Substituting $F(\beta)=\beta^{-1/2} f(\beta)$ one obtains the Bessel equation
\begin{equation}\label{eq:e12} 
\Biggl[\frac{d^2}{d\beta^2} +
\frac{1}{\beta}\frac{d}{d\beta}
 + \Biggl(k^2-\frac{\nu^2}{\beta^2}\Biggr)\Biggr]
 f(\beta) = 0,
\end{equation}
where 
\begin{equation}\label{eq:e13} 
\nu=\sqrt{\frac{L(L+1)}{3}+\frac{1}{4}}, 
\end{equation}
the boundary condition being $f(\beta_W)=0$. 
The solution of (\ref{eq:e11}), which is finite at $\beta=0$, is then 
\begin{equation}\label{eq:e14}
F(\beta)=F_{sL}(\beta) = \frac{1}{\sqrt{c}}\,\beta^{-1/2}
J_{\nu}(k_{s,\nu}\beta), 
\end{equation}
with $ k_{s,\nu}=x_{s,\nu}/\beta_W$ and $\varepsilon_{s,\nu}=k_{s,\nu}^2$,
where $x_{s,\nu}$ is the $s$-th zero of the Bessel function of the first kind 
$J_{\nu}(k_{s,\nu}\beta_W)$
and the normalization constant
$ c = \beta_W^2\,J^2_{\nu+1}(x_{s,\nu})/2$
is obtained from the condition 
$\int_{0}^{\beta_W}F_{sL}^2(\beta)\,\beta^2 d\beta = 1$.
The corresponding spectrum is then 
\begin{equation}\label{eq:e15} 
E_{s,L} = \frac{\hbar^2}{2B}\,k_{s,\nu}^2 =
\frac{\hbar^2}{2B\beta_W^2}\,x_{s,\nu}^2. 
\end{equation}
It should be noticed that in the X(5) case \cite{IacX5} 
the same Eq. (\ref{eq:e12}) occurs,
but with $\nu=\sqrt{ {L(L+1)\over 3} +{9\over 4} }$, while in the E(3) 
Euclidean algebra in 3 dimensions, which is the semidirect sum
of the T$_3$ algebra of translations in 3 dimensions and the SO(3) algebra of 
rotations in 3 dimensions \cite{Barut}, the eigenvalue equation of the square 
of the total momentum, which is a second-order Casimir operator of the 
algebra, also leads \cite{Barut,E5} to Eq. (\ref{eq:e12}), but with 
$\nu=L+{1\over 2}$.    

From the symmetry of the wave function of Eq. (\ref{eq:e8}) with respect 
to the plane which is orthogonal to the symmetry axis of the nucleus and goes 
through its center,
follows that the angular momentum $L$ can take only even nonnegative values.
Therefore no $\gamma$-bands appear in the model, as 
expected, since the $\gamma$ degree of freedom has been frozen. 

In the general case the quadrupole operator is
\begin{equation}\label{eq:e17}
T^{(E2)}_\mu = t\,\beta\Bigl[D^{2\,\ast}_{\mu,0}(\Omega)\cos\gamma +
\frac{1}{\sqrt{2}}[D^{2\,\ast}_{\mu,2}(\Omega)+D^{2\,\ast}_{\mu,-2}(\Omega)]
\sin\gamma\Bigr],
\end{equation}
where $\Omega$ denotes the Euler angles and $t$ is a scale factor.
For $\gamma=0$ the quadrupole operator becomes
\begin{equation}\label{eq:e18}
T^{(E2)}_\mu = t\,\beta\,\sqrt{\frac{4\pi}{5}}\,Y_{2\mu}(\theta,\phi).
\end{equation}

$B(E2)$ transition rates 
\begin{equation}\label{eq:e19} 
B(E2; sL \to s'L') = \frac{1}{2L+1}
\left|\langle s'L'||T^{(E2)}||sL\rangle\right|^2
\end{equation}
are calculated using the wave functions of Eq. (\ref{eq:e8}) and the volume 
element \hfill\break 
$d\tau= \beta^2 \sin\theta\, d\beta d\theta d\phi$, the final 
result being 
\begin{equation}\label{eq:e20} 
B(E2; sL \to s'L') = t^2\,\Big(C_{L\,0,\; 2\,0}^{L'\,0}\Big)^2\; 
I_{sL; s'L'}^2,
\end{equation}
where $C_{L\,0,\; 2\,0}^{L'\,0}$ are Clebsch--Gordan coefficients and 
the integrals over $\beta$ are
\begin{equation}\label{eq:e21} 
I_{sL; s'L'} = \int_{0}^{\beta_W}\beta\,F_{sL}(\beta)\,F_{s'L'}
(\beta)\,\beta^2\,d\beta. 
\end{equation}

The following remarks are now in place.

1) In both the X(3) and X(5) \cite{IacX5} models, $\gamma=0$ is considered, 
the difference being that in the former case $\gamma$ is treated as a 
parameter, while in the latter as a variable. As a consequence, separation 
of variables in X(3) is exact (because of the lack of the $\gamma$ variable), 
while in X(5) it is approximate. 

2) In both the X(3) and E(5) \cite{IacE5} models a potential depending only 
on $\beta$ is considered and exact separation of variables is achieved, 
the difference being that in the E(5) model the $\gamma$ variable remains
active, while in the X(3) case it is frozen. As a consequence, in the E(5)
case the equation involving the angles results in the solutions given 
by B\`es \cite{Bes}, while in the X(3) case the usual spherical harmonics 
occur.  

\section{Numerical results and comparison to experiment} 

The energy levels of the ground state band ($s=1$), as well as of the 
$\beta_1$ ($s=2$) and $\beta_2$ ($s=3$) bands, normalized to the energy 
of the lowest excited state, $2_1^+$, are shown in Fig.~1, together with 
intraband $B(E2)$ transition rates, normalized to the transition between 
the two lowest states, $B(E2; 2_1^+\to 0_1^+)$, while interband transitions 
are listed in Table 1.  

The energy levels of the ground state band of X(3) are also shown in Fig. 
2(a), where they are compared to the experimental data for $^{172}$Os 
\cite{172-Os} (up to the point of bandcrossing) and $^{186}$Pt \cite{186-Pt}. 
In the same figure the ground state band of X(5), along with the experimental 
data for the $N=90$ isotones $^{150}$Nd \cite{150-Nd}, $^{152}$Sm 
\cite{152-Sm}, $^{154}$Gd \cite{154-Gd}, and $^{156}$Dy \cite{156-Dy}, 
which are considered as the best realizations of X(5) 
\cite{CZ2,Kruecken,Tonev,Dewald,CaprioDy}, 
are shown for comparison. The energy levels of the $\beta_1$-band 
for the same models and nuclei are shown in Fig. 2(b), while existing 
intraband $B(E2)$ transition rates for the ground state band are shown 
in Fig. 2(c). The following comments are now in place.

1) The ground state bands of $^{172}$Os and $^{186}$Pt are in very good 
agreement with the X(3) predictions, while the $\beta_1$-bands are a little
lower. Similarly, the ground state bands of $^{150}$Nd, $^{152}$Sm, 
$^{154}$Gd, and $^{156}$Dy are in very good agreement with the X(5) 
predictions, while the $\beta_1$ bands beyond $L=4$ are much lower. 
This discrepancy is known to be fixed by considering \cite{Caprio}
a potential with linear sloped walls instead of an infinite well potential.
What occured rather unexpectedly is the fact that the $\beta_1$ bands 
of the $N=90$ isotones [the best experimental examples of X(5)] from $L=4$ 
upwards agree very well with the X(3) predictions. This could be interpreted 
as indication that the bandhead of the $\beta_1$ band is influenced by the 
presence of the $\gamma$ degree if freedom, but the excited levels of this 
band beyond $L=4$ are not influenced by it. 
Detailed measurements of intraband $B(E2)$ transition 
rates within the $\beta_1$-bands of these $N=90$ isotones could clarify
this point.  

2) Existing intraband $B(E2)$ transition rates for the ground state band
of $^{172}$Os (below the region influenced by the bandcrossing) are in good 
agreement with X(3), being quite higher than the $^{150}$Nd, $^{152}$Sm, and 
$^{154}$Gd rates, as they should. [The $B(E2)$ rates of $^{156}$Dy are known 
\cite{Dewald} to be in less good agreement with X(5), as also seen in Fig.
2(c).] However, more intraband and interband transitions (and with smaller 
error bars) are needed before 
final conclusions could be drawn. The same holds for $^{186}$Pt, for which 
experimental information on B(E2)s is missing \cite{186-Pt,McC7106}.
The relative branching ratios known in $^{186}$Pt \cite{McC7106} are 
given in Table~2, being in good agreement with the X(3) predictions.       

The placement of the above mentioned nuclei in the symmetry triangle 
\cite{Casten} of the Interacting Boson Model (IBM) \cite{IA} can be 
illuminating. All of the above mentioned N=90 isotones lie close to the phase 
coexistence and shape phase transition region of the IBM, with $^{152}$Sm 
being located on the U(5)-SU(3) side of the triangle \cite{McC67}, while 
$^{154}$Gd and $^{156}$Dy gradually move towards the center of the triangle 
\cite{McC69}. $^{172}$Os \cite{McC7105} and $^{186}$Pt \cite{McC7106} also 
appear near the center of the symmetry triangle and close to the transition 
region of the IBM. 

It should be noticed that the critical character of $^{186}$Pt is also 
supported by the criteria posed in Ref. \cite{McC726}. In particular, 
a relatively abrupt change of the $R_4=E(4_1^+)/E(2_1^+)$ ratio occurs between 
$^{186}$Pt and $^{184}$Pt, as seen in the systematics presented in 
Ref. \cite{McC7105}, while $0_2^+$ shows a minimum at $^{186}$Pt, as seen 
in the systematics presented in Ref. \cite{McC7106}, especially if the $0_2^+$
energies are normalized with respect to the $2_1^+$ state of each Pt isotope. 
Furthermore, $^{186}$Pt is located at the point where the crossover 
of $0_2^+$ and $2_\gamma^+$ occurs, as seen in the systematics presented 
in Ref. \cite{McC7106}.  

\section{Discussion}

In summary, a $\gamma$-rigid (with $\gamma=0$) version of the X(5) model 
is constructed. The model is called X(3), since it is proved that only three 
variables occur in this case, the separation of variables being exact, 
while in the X(5) case approximate separation of the five variables 
occuring there is performed. The parameter free (up to overall scale factors)
predictions of X(3) are found to be in good agreement with existing 
experimental data of $^{172}$Os and $^{186}$Pt, while a rather unexpected 
agreement of the $\beta_1$-bands of the X(5) nuclei $^{150}$Nd, $^{152}$Sm, 
$^{154}$Gd, and $^{156}$Dy to the X(3) predictions is observed. The need 
for further $B(E2)$ measurements in all of the above-mentioned nuclei 
is emphasized.   

\section*{Acknowledgements}

One of the authors (IY) is thankful to the Turkish Atomic Energy Authority 
(TAEK) for support under project number 04K120100-4.


\begin{table}
\centering
\caption{Interband $B(E2; L_i\to L_f)$ transition rates for the X(3) model,
normalized to the one between the two lowest states, $B(E2; 2_1^+\to 0_1^+)$.}
\medskip
\begin{tabular}{cr@{\qquad}cr@{\qquad}cr}
\hline\hline\noalign{\smallskip}
$L_i\to L_f$ & X(3) & $L_i\to L_f$ & X(3) & $L_i\to L_f$ & X(3) \\
\noalign{\smallskip}\hline\noalign{\smallskip}
$ 0_2 \to   2_1$ &164.0 &  & & & \\
$ 2_2 \to   4_1$ & 64.5 & $  2_2 \to   2_1$ &12.4 & $  2_2 \to   0_1$ & 0.54 \\
$ 4_2 \to   6_1$ & 42.2 & $  4_2 \to   4_1$ & 8.6 & $  4_2 \to   2_1$ & 0.43 \\
$ 6_2 \to   8_1$ & 31.1 & $  6_2 \to   6_1$ & 6.7 & $  6_2 \to   4_1$ & 0.51 \\
$ 8_2 \to  10_1$ & 24.4 & $  8_2 \to   8_1$ & 5.5 & $  8_2 \to   6_1$ & 0.56 \\
$10_2 \to  12_1$ & 19.9 & $ 10_2 \to  10_1$ & 4.7 & $ 10_2 \to   8_1$ & 0.59 \\
$12_2 \to  14_1$ & 16.6 & $ 12_2 \to  12_1$ & 4.0 & $ 12_2 \to  10_1$ & 0.60 \\
$14_2 \to  16_1$ & 14.2 & $ 14_2 \to  14_1$ & 3.5 & $ 14_2 \to  12_1$ & 0.60 \\
$16_2 \to  18_1$ & 12.3 & $ 16_2 \to  16_1$ & 3.1 & $ 16_2 \to  14_1$ & 0.60 \\
$18_2 \to  20_1$ & 10.9 & $ 18_2 \to  18_1$ & 2.8 & $ 18_2 \to  16_1$ & 0.59 \\
$20_2 \to  22_1$ &  9.7 & $ 20_2 \to  20_1$ & 2.5 & $ 20_2 \to  18_1$ & 0.58 \\
& & & & & \\
$ 0_3 \to   2_2$ &209.1 &  & & & \\
$ 2_3 \to   4_2$ & 92.0 & $  2_3 \to   2_2$ &16.2 & $  2_3 \to   0_2$ & 0.67 \\
$ 4_3 \to   6_2$ & 65.3 & $  4_3 \to   4_2$ &12.2 & $  4_3 \to   2_2$ & 0.47 \\
$ 6_3 \to   8_2$ & 50.9 & $  6_3 \to   6_2$ &10.1 & $  6_3 \to   4_2$ & 0.52 \\
$ 8_3 \to  10_2$ & 41.6 & $  8_3 \to   8_2$ & 8.6 & $  8_3 \to   6_2$ & 0.57 \\
$10_3 \to  12_2$ & 35.0 & $ 10_3 \to  10_2$ & 7.5 & $ 10_3 \to   8_2$ & 0.61 \\
$12_3 \to  14_2$ & 30.1 & $ 12_3 \to  12_2$ & 6.6 & $ 12_3 \to  10_2$ & 0.63 \\
$14_3 \to  16_2$ & 26.3 & $ 14_3 \to  14_2$ & 5.9 & $ 14_3 \to  12_2$ & 0.65 \\
$16_3 \to  18_2$ & 23.3 & $ 16_3 \to  16_2$ & 5.4 & $ 16_3 \to  14_2$ & 0.66 \\
$18_3 \to  20_2$ & 20.8 & $ 18_3 \to  18_2$ & 4.9 & $ 18_3 \to  16_2$ & 0.66 \\
$20_3 \to  22_2$ & 18.8 & $ 20_3 \to  20_2$ & 4.5 & $ 20_3 \to  18_2$ & 0.66 \\
\noalign{\smallskip}\hline\hline
\end{tabular}
\end{table}


\begin{table}
\centering
\caption{Relative $B(E2)$ branching ratios for the X(3) model compared 
to existing exprerimental data \cite{McC7106} for $^{186}$Pt.}  
\medskip
\begin{tabular}{cr@{\qquad}cr@{\qquad}cr}
\hline\hline\noalign{\smallskip}
$L_i\to L_f$ & exp. & X(3) & $L_i\to L_f$ & exp. & X(3) \\
\noalign{\smallskip}\hline\noalign{\smallskip}
$2_2\to 0_2$ & 100 & 100  & $4_2\to 2_2$ & 100 & 100 \\
$2_2\to 0_1$ & 8(1) & 0.7 & $4_2\to 2_1$ & 2.6(3) & 0.3 \\
$2_2\to 4_1$ & 68(7) & 80 & $4_2\to 4_1$ & $<12$ & 6 \\
\noalign{\smallskip}\hline\hline
\end{tabular}
\end{table}


\begin{figure}[ht]
\center{\includegraphics[height=180mm]{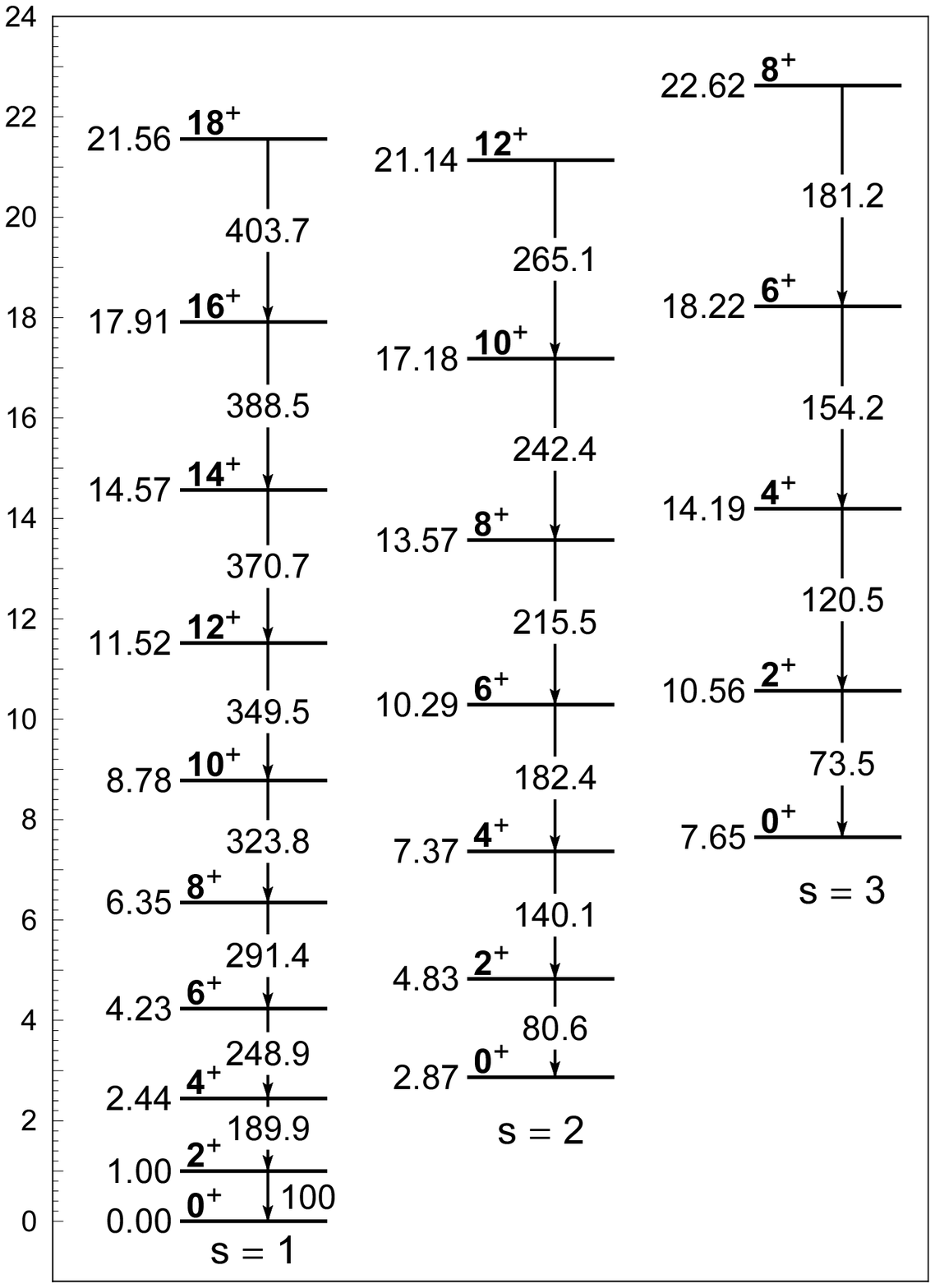}} 
\caption{
 Energy levels of the ground state ($s=1$), $\beta_1$ ($s=2$), and $\beta_2$ 
($s=3$) bands of X(3), normalized to the energy 
of the lowest excited state, $2_1^+$, together with 
intraband $B(E2)$ transition rates, normalized to the transition between 
the two lowest states, $B(E2; 2_1^+\to 0_1^+)$. Interband transitions 
are listed in Table 1.  See Section 3 for further discussion. }
\end{figure}


\begin{figure}[ht]
\rotatebox{270}{\includegraphics[height=80mm]{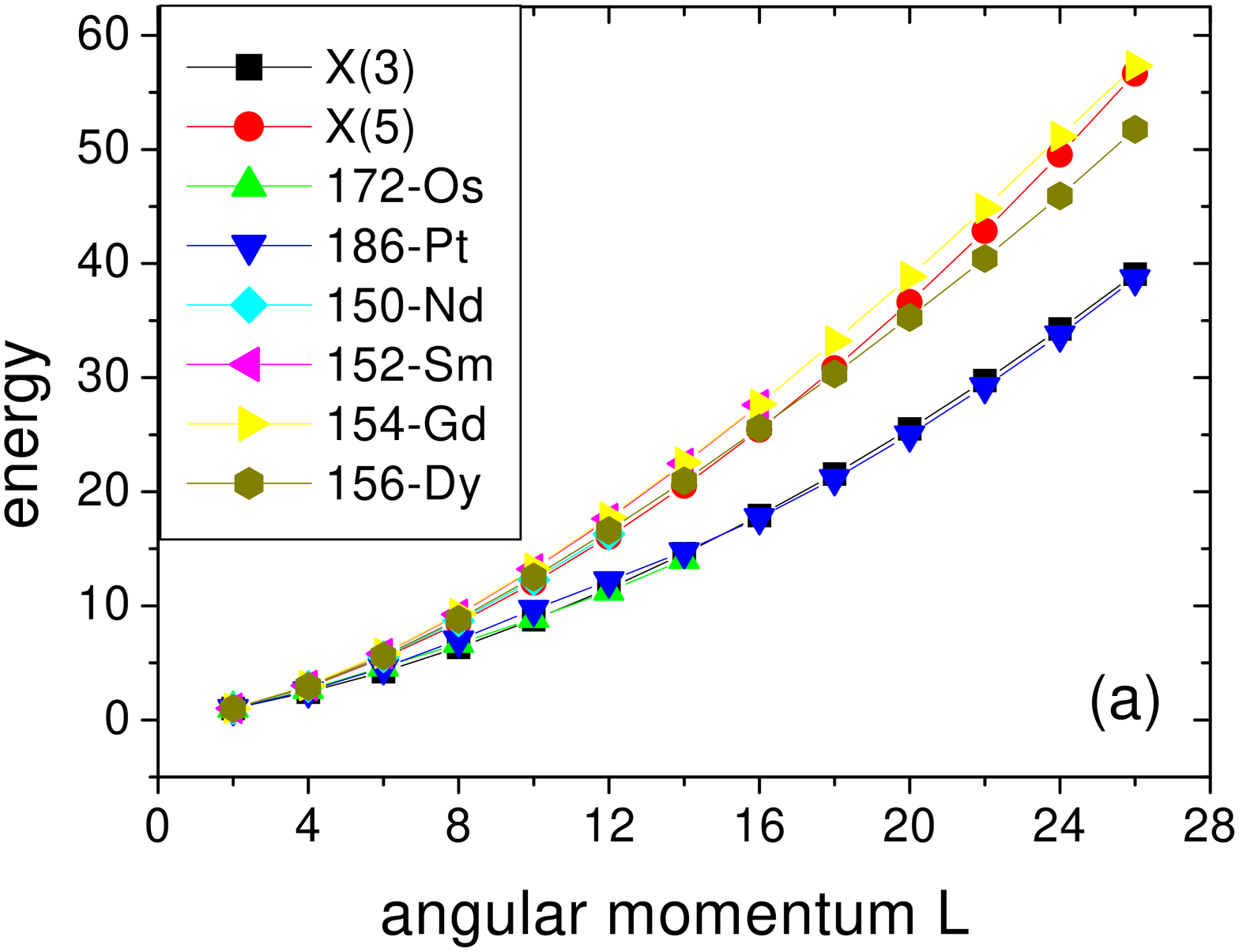}} 
\rotatebox{270}{\includegraphics[height=80mm]{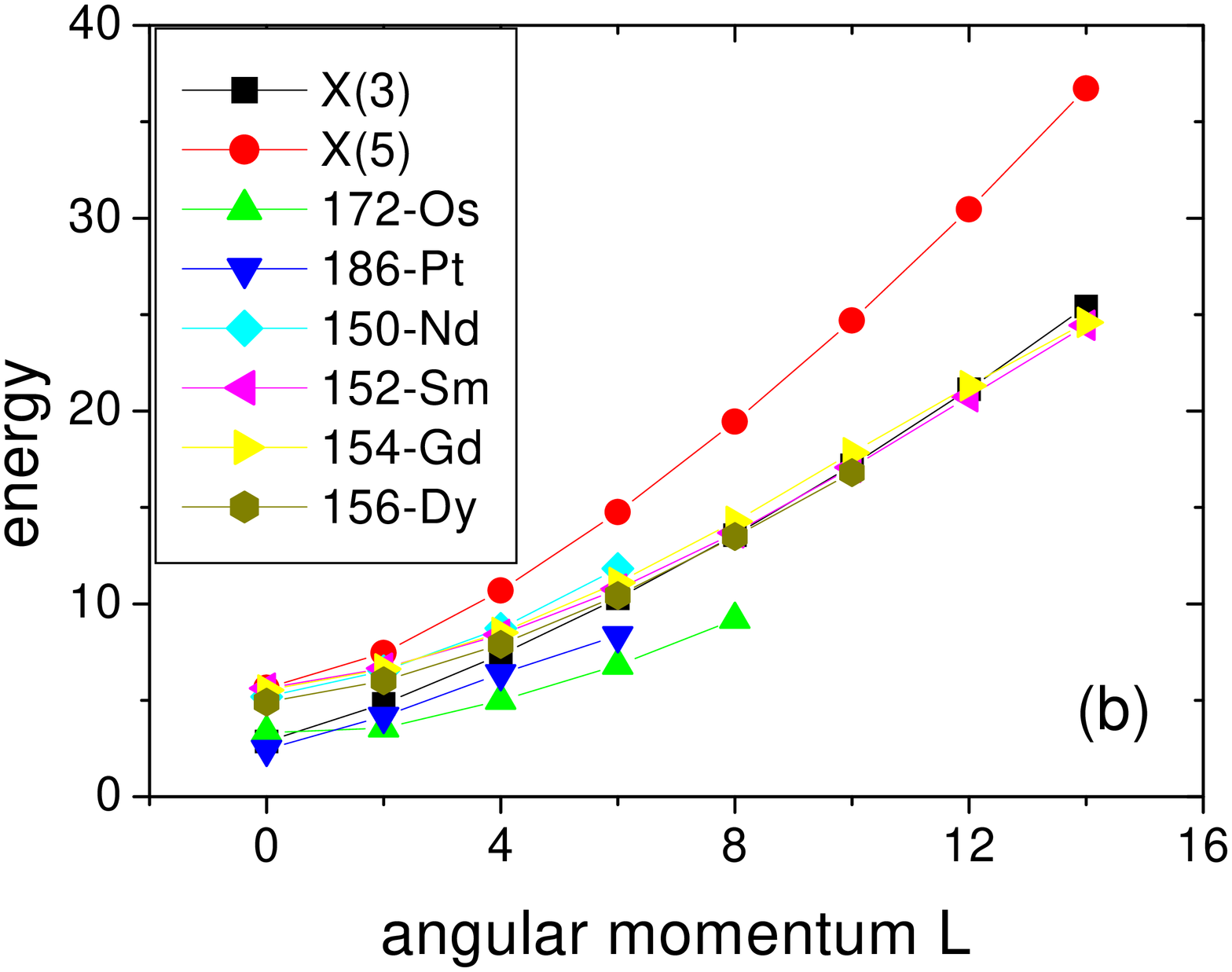}} 
\rotatebox{270}{\includegraphics[height=80mm]{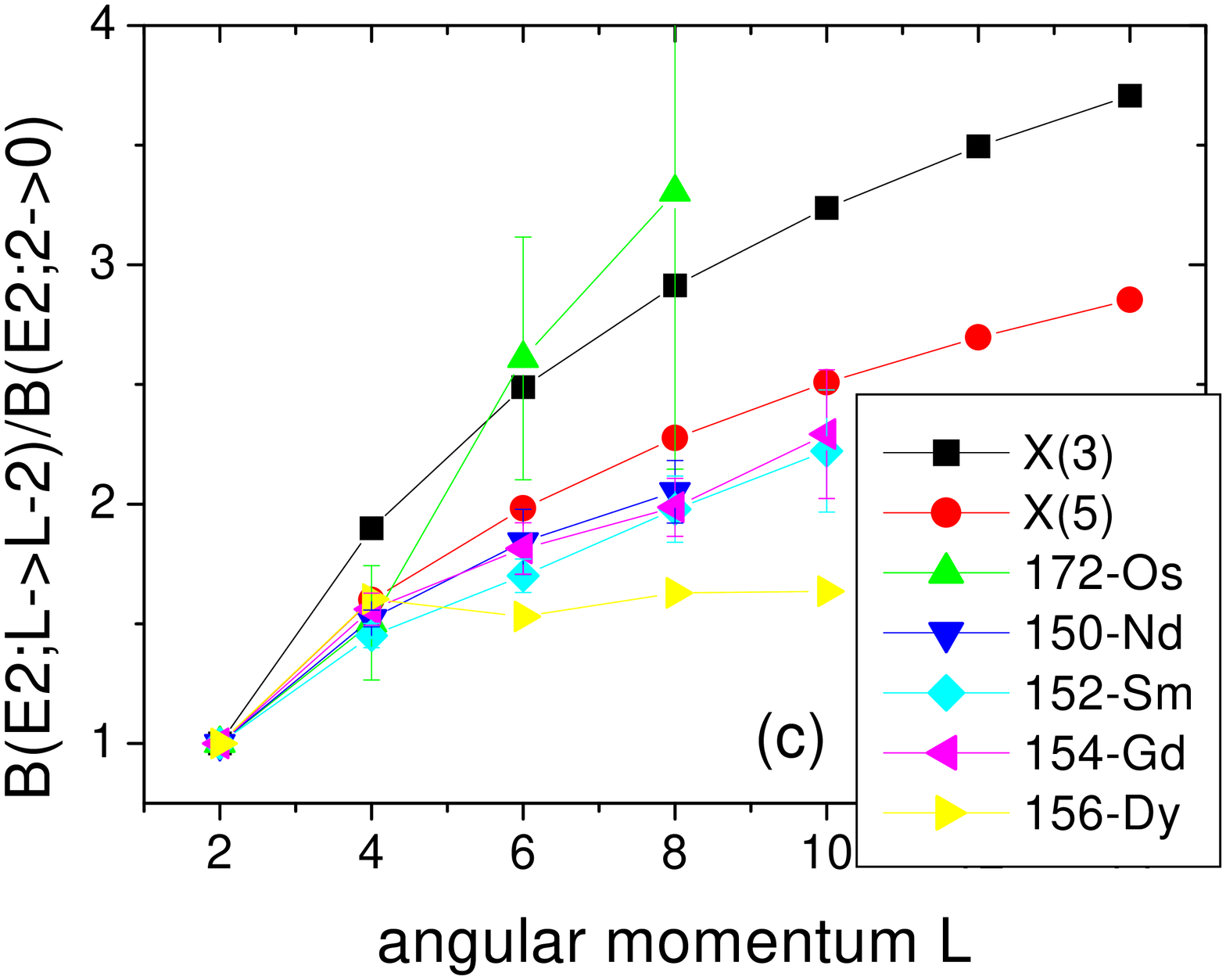}} 
\caption{
(a) Energy levels of the ground state bands of the X(3) and X(5) \cite{IacX5} 
models, compared to experimental data for $^{172}$Os \cite{172-Os}, 
$^{186}$Pt \cite{186-Pt}, $^{150}$Nd \cite{150-Nd}, $^{152}$Sm \cite{152-Sm}, 
$^{154}$Gd \cite{154-Gd}, and $^{156}$Dy \cite{156-Dy}. The levels of each 
band are normalized to the $2_1^+$ state. 
(b) Same for the $\beta_1$-bands, also normalized to the $2_1^+$ state. 
(c) Same for existing intraband $B(E2)$ transition rates within the ground 
state band, normalized to the $B(E2; 2_1^+\to 0_1^+)$ rate. The data for 
$^{156}$Dy are taken from Ref. \cite{Dewald}. See Section 3 for further 
discussion.}
\end{figure}  


\newpage

\begin{thebibliography}{99}

\bibitem{IacE5}
F. Iachello, Phys. Rev. Lett.  85 (2000) 3580. 

\bibitem{IacX5}
F. Iachello, Phys. Rev. Lett.  87 (2001) 052502. 

\bibitem{CZ1}
R. F. Casten, N. V. Zamfir, Phys. Rev. Lett.  85 (2000) 3584. 

\bibitem{Clark1}
R. M. Clark, et al., Phys. Rev. C 69 (2004) 064322. 

\bibitem{CZ2} 
R. F. Casten, N. V. Zamfir, Phys. Rev. Lett.  87 (2001) 052503. 

\bibitem{Clark2}
R. M. Clark, et al., Phys. Rev. C 68 (2003) 037301. 

\bibitem{Kruecken} 
R. Kr\"ucken, et al., Phys. Rev. Lett. 88 (2002) 232501. 

\bibitem{Tonev} 
D. Tonev, et al., Phys. Rev. C 69 (2004) 034334. 

\bibitem{Dewald}
A. Dewald, et al., Eur. Phys. J. A  20 (2004) 173. 

\bibitem{CaprioDy}
M. A. Caprio, et al., Phys. Rev. C  66 (2002) 054310.  

\bibitem{Bohr}
A. Bohr, Mat. Fys. Medd. K. Dan. Vidensk. Selsk.  26, no. 14 (1952).

\bibitem{DC}
A. S. Davydov, A. A. Chaban, Nucl. Phys. 20 (1960) 499.

\bibitem{ST}
A. G. Sitenko, V. K. Tartakovskii, Lectures on the Theory of the 
Nucleus, Atomizdat, Moscow, 1972 (in Russian).

\bibitem{Zare}
R. N. Zare, Angular Momentum, Wiley, New York, 1988.

\bibitem{Davydov}
A. S. Davydov, Theory of the Atomic Nucleus, Fizmatgiz, Moscow, 1958. 

\bibitem{EG}
J. M. Eisenberg, W. Greiner, Nuclear Theory, Vol. I: Nuclear 
Models, North-Holland, Amsterdam, 1970.

\bibitem{Barut}
A. O. Barut, R Raczka, Theory of Group Representations and 
Applications, World Scientific, Singapore, 1986. 

\bibitem{E5}
D. Bonatsos, D. Lenis, N. Minkov, P. P. Raychev, P. A. Terziev, 
Phys. Rev. C 69 (2004) 044316. 

\bibitem{Bes}
D. R. B\`es, Nucl. Phys. 10 (1959) 373. 

\bibitem{172-Os}
B. Singh, Nucl. Data Sheets 75 (1995) 199. 

\bibitem{186-Pt}
C. M. Baglin, Nucl. Data Sheets 99 (2003) 1. 

\bibitem{150-Nd}
E. der Mateosian, J. K. Tuli, Nucl. Data Sheets 75 (1995) 827. 

\bibitem{152-Sm}
A. Artna-Cohen, Nucl. Data Sheets 79 (1996) 1. 

\bibitem{154-Gd}
C. W. Reich, R. G. Helmer, Nucl. Data Sheets 85 (1998) 171. 

\bibitem{156-Dy}
C. W. Reich, Nucl. Data Sheets 99 (2003) 753. 

\bibitem{Caprio}
M. A. Caprio, Phys. Rev. C 69 (2004) 044307. 

\bibitem{McC7106}
E. A. McCutchan, R. F. Casten, and N. V. Zamfir, Phys. Rev. C 71 (2005) 
061301. 

\bibitem{Casten}
R. F. Casten, Nuclear Structure from a Simple Perspective, Oxford University 
Press, Oxford, 1990. 

\bibitem{IA}
F. Iachello, A. Arima, The Interacting Boson Model, Cambridge 
University Press, Cambridge, 1987. 

\bibitem{McC67}
N. V. Zamfir, E. A. McCutchan, and R. F. Casten, Yad. Fiz. 67 (2004) 1856
[Phys. At. Nucl. 67 (2004) 1829]. 

\bibitem{McC69}
E. A. McCutchan, N. V. Zamfir, and R. F. Casten, Phys. Rev. C 69 (2004) 
063406. 

\bibitem{McC7105}
E. A. McCutchan and N. V. Zamfir, Phys. Rev. C 71 (2005) 054306. 

\bibitem{McC726}
N. V. Zamfir, E. A. McCutchan, and R. F. Casten, in Nuclear Physics, Large 
and Small: International Conference on Microscopic Studies of Collective 
Phenomena, ed. R. Bijker, R. F. Casten, and A. Frank, AIP Conf. Proc. 
726 (2004) 187. 

\end{thebibliography}
\end{document}